\documentstyle[aps]{revtex}

\begin{document}
\title{Two-object remote quantum control}
\author{Yafei Yu$^{\thanks{%
corresponding author, e-mail: yfyu@wipm.ac.cn }a}$, Tangkun Liu$^{a,b}$
,Yanxia Huang $^{a,b}$and Mingsheng Zhan $^{a}$}
\address{$^{a}$State Key Laboratory of Magnetic Resonance and Atomic and Molecular\\
Physics, Wuhan Institute of Physics and Mathematics, Chinese Academy of\\
Sciences, Wuhan 430071, PR China\\
$^{b}$Department of Physics, Hubei Normal University, Huangshi 435000,\\
PR China}
\maketitle

\begin{abstract}
We consider the two-object remote quantum control for a special case in
which all the object qubits are in a telecloning state. We propose a scheme
which achieves the two-object remote quantum control by using two particular
four-particle entangled states.

PACS number(s): 03.67.-a, 03.65.Bz

\end{abstract}

Quantum entanglement exhibits nonlocal correlation between separate systems,
which describes that the results of measurements on one system can not be
specified independent of the parameters of the measurements on the other
systems. The quantum entanglement in multipartite systems describes the
quantum correlation between more than two parties, and is more complex and
richer than the bipartite entanglement. For example, the bound entanglement 
\cite{63/4/5} is a distinguished kind of entanglement in multipartite
systems. The multipartite quantum entanglement has been extensively applied
into quantum information processing and quantum computation, such as quantum
telecloning \cite{2}, quantum remote information concentration\cite{3} and
multi-output programmable quantum processor \cite{4}. Here we apply the
multipartite quantum entanglement into remote quantum control \cite{5}.
Recently quantum remote control is deeply studied by S. F. Huelga \cite{5,6}%
, B. Reznik \cite{7} and Chui-ping Yang \cite{8} et al. The previous works
are based on one object or aimed to creating interaction between separate
systems. However, it is important to remotely control several objects
simultaneously and individually in quantum network. For simplicity, we
consider two-object remote control.

An unknown state $\left| \chi \right\rangle =\alpha \left| 0\right\rangle
+\beta \left| 1\right\rangle $ of a single qubit can be teleported to two
spatially separated receivers by a telecloning process\cite{2}
simultaneously. After the telecloning process, the qubits held by the two
receivers and an acillary qubit are in a pure three-qubit state $\left| \psi
\right\rangle _{ABC}=\alpha \left| \phi _{0}\right\rangle _{ABC}+\beta
\left| \phi _{1}\right\rangle _{ABC}$, where $\alpha $ and $\beta $ are real
numbers, and satisfy $\alpha ^{2}+\beta ^{2}=1$. $\left| \phi
_{0}\right\rangle _{ABC}$ and $\left| \phi _{1}\right\rangle _{ABC}$ are
defined as: 
\begin{eqnarray}
\left| \phi _{0}\right\rangle _{ABC} &=&\sqrt{\frac{2}{3}}\left|
0\right\rangle _{A}\left| 0\right\rangle _{B}\left| 0\right\rangle _{C}+%
\sqrt{\frac{1}{6}}(\left| 1\right\rangle _{A}\left| 0\right\rangle
_{B}\left| 1\right\rangle _{C}+\left| 1\right\rangle _{A}\left|
1\right\rangle _{B}\left| 1\right\rangle _{C}), \\
\left| \phi _{1}\right\rangle _{ABC} &=&\sqrt{\frac{2}{3}}\left|
1\right\rangle _{A}\left| 1\right\rangle _{B}\left| 1\right\rangle _{C}+%
\sqrt{\frac{1}{6}}(\left| 0\right\rangle _{A}\left| 0\right\rangle
_{B}\left| 1\right\rangle _{C}+\left| 0\right\rangle _{A}\left|
1\right\rangle _{B}\left| 0\right\rangle _{C}),  \nonumber
\end{eqnarray}
where the qubits B and C are held by the two receivers, the qubit A is an
ancilla. That is, the two receivers obtain an optimal copy of the state $%
\left| \chi \right\rangle $ with the fidelity of $\frac{5}{6}$,
respectively. In terms of the density matrices, namely, 
\begin{eqnarray}
\rho _{B} &=&Tr_{A,C}(\left| \psi \right\rangle _{ABC}\left\langle \psi
\right| )=\frac{5}{6}\left| \chi \right\rangle \left\langle \chi \right| +%
\frac{1}{6}\left| \chi ^{\bot }\right\rangle \left\langle \chi ^{\perp
}\right| ,  \nonumber \\
\rho _{C} &=&Tr_{A,B}(\left| \psi \right\rangle _{ABC}\left\langle \psi
\right| )=\frac{5}{6}\left| \chi \right\rangle \left\langle \chi \right| +%
\frac{1}{6}\left| \chi ^{\bot }\right\rangle \left\langle \chi ^{\perp
}\right| ,
\end{eqnarray}
where $\left| \chi ^{\bot }\right\rangle $ denotes a state orthogonal to the
state $\left| \chi \right\rangle $. Now the problem present is that if the
sender is required to remotely control the two optimal cloning state only by
local operations and classical communication (LOCC), how do the sender, i.e.
the controller to do it? Namely, the controller will remotely control two
objects (the two optimal cloning state) by LOCC simultaneously. The task of
the two-object remote control can be achieved by pre-sharing two particular
four-partite entangled states, a four-partite unlockable bound entangled
state\cite{63} 
\[
\rho _{DEFG}^{ub}=\frac{1}{4}\sum_{i=0}^{3}\left| \Phi ^{i}\right\rangle
_{DE}\left\langle \Phi ^{i}\right| \otimes \left| \Phi ^{i}\right\rangle
_{FG}\left\langle \Phi ^{i}\right| , 
\]
where $\left| \Phi ^{i}\right\rangle _{(i=0,1,2,3)}$ represent the four Bell
states $\left| \Phi ^{0,1}\right\rangle =\left( \left| 00\right\rangle \pm
\left| 11\right\rangle \right) /\sqrt{2}$, $\left| \Phi ^{2,3}\right\rangle
=\left( \left| 01\right\rangle \pm \left| 01\right\rangle \right) /\sqrt{2}$%
, and a four-particle free entangled state 
\begin{equation}
\left| \varphi \right\rangle _{PA^{^{\prime }}B^{^{\prime }}C^{^{\prime }}}=%
\frac{1}{\sqrt{2}}\left( \left| 0\right\rangle _{P}\left| \phi
_{0}\right\rangle _{A^{^{\prime }}B^{^{\prime }}C^{^{\prime }}}+\left|
1\right\rangle _{P}\left| \phi _{1}\right\rangle _{A^{^{\prime }}B^{^{\prime
}}C^{^{\prime }}}\right) ,
\end{equation}
where the states $\left| \phi _{0}\right\rangle _{A^{^{\prime }}B^{^{\prime
}}C^{^{\prime }}}$ and $\left| \phi _{1}\right\rangle _{A^{^{\prime
}}B^{^{\prime }}C^{^{\prime }}}$ are defined as in Eq. (1). The procedure is
illuminated in the Fig. 1.

Supposing that three spatially separated parties Alice, Bob and Charlie hold
qubits A, B and C in the state $\left| \psi \right\rangle _{ABC}$,
respectively. The controller sends in advance the qubits E and A$^{^{\prime
}}$ to Alice, the qubits F and B$^{^{\prime }}$ to Bob, and the qubits G and
C$^{^{\prime }}$ to Charlie, and leaves the qubits D and P for himself. Then
the controller begins the work of two-object remote control by LOCC.

First, by exploiting the bound entangled state $\rho _{DEFG}^{ub}$, the
controller concentrates the information diluted in the state $\left| \psi
\right\rangle _{ABC}$ back to the state $\left| \chi \right\rangle $ of the
qubit D under the cooperations of the three parties Alice, Bob and Charlie%
\cite{3}. The three parties are asked to perform individually the Bell-state
measurements (BSMs) on the pairs of the qubtis in hand, i.e. the qubits A
and E, the qubits B and F, and the qubits C and G. Each of them obtains one
of the possible outputs $\left\{ \left| \Phi ^{i}\right\rangle
_{(i=0,1,2,3)}\right\} $ of the BSM, which is associated with a
corresponding Pauli operator in the set $\left\{ \sigma
_{(i=0,1,2,3)}^{i}\right\} $, and communicates the result with the
controller, respectively. The controller determines a Pauli operator $\sigma
^{i}$ on his qubit D for retrieving the state $\left| \chi \right\rangle $
on the qubit D, according the product of the three Pauli operators
pertaining to each one of the three BSMs. That is, up a global phase factor, 
$\sigma ^{i}$ is equal to $\sigma _{AE}^{l}\sigma _{BF}^{j}\sigma _{CG}^{k}$%
, where $l,j,k=0,1,2,3$ and the subscripts denote the BSMs on the
corresponding pairs of qubits. Finally, the pure state $\left| \chi
\right\rangle $ is recreated on the qubit D in the location of the sender
(i.e. the controller).

The process can be analytically interpreted. We joint the state $\left| \psi
\right\rangle _{ABC}$ with $\rho _{DEFG}^{ub}$: 
\begin{eqnarray}
\left| \psi \right\rangle _{ABC}\otimes \rho _{DEFG}^{ub}\otimes
_{ABC}\left\langle \psi \right| &=&(\alpha \left| \phi _{0}\right\rangle
_{ABC}+\beta \left| \phi _{1}\right\rangle _{ABC})\rho _{DEFG}^{ub}(\alpha
_{ABC}^{*}\left\langle \phi _{0}\right| +\beta _{ABC}^{*}\left\langle \phi
_{1}\right| )  \nonumber \\
&\rightarrow &\frac{1}{64}\{(\left| I\right\rangle \left\langle I\right|
)_{AE,BF,CG}\sigma _{0}(\alpha \left| 0\right\rangle _{D}+\beta \left|
1\right\rangle _{D})(\alpha _{D}^{*}\left\langle 0\right| +\beta
_{D}^{*}\left\langle 1\right| )\sigma _{0}  \nonumber \\
&&+(\left| II\right\rangle \left\langle II\right| )_{AE,BF,CG}\sigma
_{1}(\alpha \left| 0\right\rangle _{D}+\beta \left| 1\right\rangle
_{D})(\alpha _{D}^{*}\left\langle 0\right| +\beta _{D}^{*}\left\langle
1\right| )\sigma _{1}  \nonumber \\
&&+(\left| III\right\rangle \left\langle III\right| )_{AE,BF,CG}\sigma
_{2}(\alpha \left| 0\right\rangle _{D}+\beta \left| 1\right\rangle
_{D})(\alpha _{D}^{*}\left\langle 0\right| +\beta _{D}^{*}\left\langle
1\right| )\sigma _{2}  \nonumber \\
&&+(\left| IV\right\rangle \left\langle IV\right| )_{AE,BF,CG}\sigma
_{3}(\alpha \left| 0\right\rangle _{D}+\beta \left| 1\right\rangle
_{D})(\alpha _{D}^{*}\left\langle 0\right| +\beta _{D}^{*}\left\langle
1\right| )\sigma _{3}\},
\end{eqnarray}
where$\left| I\right\rangle \left\langle I\right| $ marks a set of the
combining results of the three BSMs as 
\begin{eqnarray}
\left| I\right\rangle \left\langle I\right| &=&\left| \Phi ^{0}\right\rangle
_{AE}\left\langle \Phi ^{0}\right| \otimes \left| \Phi ^{0}\right\rangle
_{BF}\left\langle \Phi ^{0}\right| \otimes \left| \Phi ^{0}\right\rangle
_{CG}\left\langle \Phi ^{0}\right| +\left| \Phi ^{0}\right\rangle
_{AE}\left\langle \Phi ^{0}\right| \otimes \left| \Phi ^{1}\right\rangle
_{BF}\left\langle \Phi ^{1}\right| \otimes \left| \Phi ^{1}\right\rangle
_{CG}\left\langle \Phi ^{1}\right|  \nonumber \\
&&+\left| \Phi ^{1}\right\rangle _{AE}\left\langle \Phi ^{1}\right| \otimes
\left| \Phi ^{0}\right\rangle _{BF}\left\langle \Phi ^{0}\right| \otimes
\left| \Phi ^{1}\right\rangle _{CG}\left\langle \Phi ^{1}\right| +\left|
\Phi ^{1}\right\rangle _{AE}\left\langle \Phi ^{1}\right| \otimes \left|
\Phi ^{1}\right\rangle _{BF}\left\langle \Phi ^{1}\right| \otimes \left|
\Phi ^{0}\right\rangle _{CG}\left\langle \Phi ^{0}\right|  \nonumber \\
&&+\left| \Phi ^{2}\right\rangle _{AE}\left\langle \Phi ^{2}\right| \otimes
\left| \Phi ^{2}\right\rangle _{BF}\left\langle \Phi ^{2}\right| \otimes
\left| \Phi ^{0}\right\rangle _{CG}\left\langle \Phi ^{0}\right| +\left|
\Phi ^{2}\right\rangle _{AE}\left\langle \Phi ^{2}\right| \otimes \left|
\Phi ^{3}\right\rangle _{BF}\left\langle \Phi ^{3}\right| \otimes \left|
\Phi ^{1}\right\rangle _{CG}\left\langle \Phi ^{1}\right|  \nonumber \\
&&+\left| \Phi ^{3}\right\rangle _{AE}\left\langle \Phi ^{3}\right| \otimes
\left| \Phi ^{2}\right\rangle _{BF}\left\langle \Phi ^{2}\right| \otimes
\left| \Phi ^{1}\right\rangle _{CG}\left\langle \Phi ^{1}\right| +\left|
\Phi ^{3}\right\rangle _{AE}\left\langle \Phi ^{3}\right| \otimes \left|
\Phi ^{3}\right\rangle _{BF}\left\langle \Phi ^{3}\right| \otimes \left|
\Phi ^{0}\right\rangle _{CG}\left\langle \Phi ^{0}\right|  \nonumber \\
&&+\left| \Phi ^{0}\right\rangle _{AE}\left\langle \Phi ^{0}\right| \otimes
\left| \Phi ^{2}\right\rangle _{BF}\left\langle \Phi ^{2}\right| \otimes
\left| \Phi ^{2}\right\rangle _{CG}\left\langle \Phi ^{2}\right| +\left|
\Phi ^{0}\right\rangle _{AE}\left\langle \Phi ^{0}\right| \otimes \left|
\Phi ^{3}\right\rangle _{BF}\left\langle \Phi ^{3}\right| \otimes \left|
\Phi ^{3}\right\rangle _{CG}\left\langle \Phi ^{3}\right|  \nonumber \\
&&+\left| \Phi ^{1}\right\rangle _{AE}\left\langle \Phi ^{1}\right| \otimes
\left| \Phi ^{2}\right\rangle _{BF}\left\langle \Phi ^{2}\right| \otimes
\left| \Phi ^{3}\right\rangle _{CG}\left\langle \Phi ^{3}\right| +\left|
\Phi ^{1}\right\rangle _{AE}\left\langle \Phi ^{1}\right| \otimes \left|
\Phi ^{3}\right\rangle _{BF}\left\langle \Phi ^{3}\right| \otimes \left|
\Phi ^{2}\right\rangle _{CG}\left\langle \Phi ^{2}\right|  \nonumber \\
&&+\left| \Phi ^{2}\right\rangle _{AE}\left\langle \Phi ^{2}\right| \otimes
\left| \Phi ^{0}\right\rangle _{BF}\left\langle \Phi ^{0}\right| \otimes
\left| \Phi ^{2}\right\rangle _{CG}\left\langle \Phi ^{2}\right| +\left|
\Phi ^{2}\right\rangle _{AE}\left\langle \Phi ^{2}\right| \otimes \left|
\Phi ^{1}\right\rangle _{BF}\left\langle \Phi ^{1}\right| \otimes \left|
\Phi ^{3}\right\rangle _{CG}\left\langle \Phi ^{3}\right|  \nonumber \\
&&+\left| \Phi ^{3}\right\rangle _{AE}\left\langle \Phi ^{3}\right| \otimes
\left| \Phi ^{0}\right\rangle _{BF}\left\langle \Phi ^{0}\right| \otimes
\left| \Phi ^{3}\right\rangle _{CG}\left\langle \Phi ^{3}\right| +\left|
\Phi ^{3}\right\rangle _{AE}\left\langle \Phi ^{3}\right| \otimes \left|
\Phi ^{1}\right\rangle _{BF}\left\langle \Phi ^{1}\right| \otimes \left|
\Phi ^{2}\right\rangle _{CG}\left\langle \Phi ^{2}\right| .
\end{eqnarray}
It is clear that the combing results of the three BSMs $\left|
I\right\rangle \left\langle I\right| $ correspond to the case that the
product $\sigma _{AE}^{l}\sigma _{BF}^{j}\sigma _{CG(l,j,k=0,1,2,3)}^{k}$ of
the three Pauli operators pertaining to each one of the three BSMs amounts
to $\sigma ^{0}$. So the state of the qubit D is projected to the state $%
\left| \chi \right\rangle $. The rest may be deduced by analogy that $\left|
II\right\rangle \left\langle II\right| $ corresponds to the product of $%
\sigma ^{1}$, $\left| III\right\rangle \left\langle III\right| $ to the
product of $\sigma ^{2}$, $\left| IV\right\rangle \left\langle IV\right| $
to the product of $\sigma ^{3}$. Correspondingly, the controller does a
Pauli operator $\sigma ^{1}$, $\sigma ^{2}$ or $\sigma ^{3}$ on his qubit D
in order to get the correct state $\left| \chi \right\rangle $,
respectively. In the end, the controller gets a qubit D in the state $\left|
\chi \right\rangle $ with certainty.

Secondly, the controller manipulates the qubit D, transforms the state of
the qubit D unitarily, and gets a state $\left| \chi ^{^{\prime
}}\right\rangle _{D}=U\left| \chi \right\rangle _{D}=\alpha ^{^{\prime
}}\left| 0\right\rangle _{D}+\beta ^{^{\prime }}e^{i\theta }\left|
1\right\rangle _{D}$, where $\alpha ^{^{\prime }}$ and $\beta ^{^{\prime }}$
are real numbers, and satisfy $\alpha ^{^{\prime }2}+\beta ^{^{\prime }2}=1$%
, $\theta \in (0,2\pi ]$.

Finally, under the aid of the free entangled state $\left| \varphi
\right\rangle _{PA^{^{\prime }}B^{^{\prime }}C^{^{\prime }}}$, the
controller again distributs the transformed state $\left| \chi ^{^{\prime
}}\right\rangle _{D}$ to the parties Alice, Bob and Charlie by a telecloning
process. The controller performs the Bell-state measurement on the qubits D
and P, and obtains one of the four possible outcomes $\left\{ \left| \Phi
^{i}\right\rangle _{(i=0,1,2,3)}\right\} $ of the Bell-state measurement.
Then he broadcasts the results to the three parties Alice, Bob and Charlie.
The three parties rotate their respective qubits A$^{^{\prime }}$, B$%
^{^{\prime }}$ and C$^{^{\prime }}$ by a corresponding Pauli operator $%
\sigma ^{i}$ pertaining to the the result of the BSM. At last, the state of
the three qubits A$^{^{\prime }}$, B$^{^{\prime }}$ and C$^{^{\prime }}$ are
mapped to a state 
\begin{equation}
\left| \psi ^{^{\prime }}\right\rangle _{A^{^{\prime }}B^{^{\prime
}}C^{^{\prime }}}=\alpha ^{^{\prime }}\left| \phi _{0}\right\rangle
_{A^{^{\prime }}B^{^{\prime }}C^{^{\prime }}}+\beta ^{^{\prime }}e^{i\theta
}\left| \phi _{1}\right\rangle _{A^{^{\prime }}B^{^{\prime }}C^{^{\prime }}}
\end{equation}
with respect to the state $\left| \chi ^{^{\prime }}\right\rangle _{D}$. The
process also can be simply expressed in a formular way, 
\begin{eqnarray*}
\left| \chi ^{^{\prime }}\right\rangle _{D}\otimes \left| \varphi
\right\rangle _{PA^{^{\prime }}B^{^{\prime }}C^{^{\prime }}} &=&(\alpha
^{/}\left| 0\right\rangle _{D}+\beta ^{/}e^{i\theta }\left| 1\right\rangle
_{D})\otimes \frac{1}{\sqrt{2}}\left( \left| 0\right\rangle _{P}\left| \phi
_{0}\right\rangle _{A^{^{\prime }}B^{^{\prime }}C^{^{\prime }}}+\left|
1\right\rangle _{P}\left| \phi _{1}\right\rangle _{A^{^{\prime }}B^{^{\prime
}}C^{^{\prime }}}\right) \\
&=&\frac{1}{2}\{\left| \Phi ^{0}\right\rangle _{DP}(\alpha ^{^{\prime
}}\left| \phi _{0}\right\rangle _{A^{^{\prime }}B^{^{\prime }}C^{^{\prime
}}}+\beta ^{^{\prime }}e^{i\theta }\left| \phi _{1}\right\rangle
_{A^{^{\prime }}B^{^{\prime }}C^{^{\prime }}}) \\
&&+\left| \Phi ^{1}\right\rangle _{DP}(\alpha ^{^{\prime }}\left| \phi
_{0}\right\rangle _{A^{^{\prime }}B^{^{\prime }}C^{^{\prime }}}-\beta
^{^{\prime }}e^{i\theta }\left| \phi _{1}\right\rangle _{A^{^{\prime
}}B^{^{\prime }}C^{^{\prime }}}) \\
&&+\left| \Phi ^{2}\right\rangle _{DP}(\alpha ^{^{\prime }}\left| \phi
_{1}\right\rangle _{A^{^{\prime }}B^{^{\prime }}C^{^{\prime }}}+\beta
^{^{\prime }}e^{i\theta }\left| \phi _{0}\right\rangle _{A^{^{\prime
}}B^{^{\prime }}C^{^{\prime }}}) \\
&&+\left| \Phi ^{3}\right\rangle _{DP}(\alpha ^{^{\prime }}\left| \phi
_{1}\right\rangle _{A^{^{\prime }}B^{^{\prime }}C^{^{\prime }}}-\beta
^{^{\prime }}e^{i\theta }\left| \phi _{0}\right\rangle _{A^{^{\prime
}}B^{^{\prime }}C^{^{\prime }}})\}.
\end{eqnarray*}
It is shown that the state $\left| \psi ^{^{\prime }}\right\rangle
_{A^{^{\prime }}B^{^{\prime }}C^{^{\prime }}}=\alpha ^{^{\prime }}\left|
\phi _{0}\right\rangle _{A^{^{\prime }}B^{^{\prime }}C^{^{\prime }}}+\beta
^{^{\prime }}e^{i\theta }\left| \phi _{1}\right\rangle _{A^{^{\prime
}}B^{^{\prime }}C^{^{\prime }}}$ can be retrived by performing corresponding
Pauli operators $\sigma _{(i=0,1,2,3)}^{i}$ on each of the three qubits A$%
^{^{\prime }}$, B$^{^{\prime }}$ and C$^{^{\prime }}$ as analyzed in \cite{2}%
.

From the state $\left| \psi ^{^{\prime }}\right\rangle _{A^{^{\prime
}}B^{^{\prime }}C^{^{\prime }}}$, after tracing out ancillary qubit A$%
^{^{\prime }}$ and the other qubit the density matrices of the qubits B$%
^{^{\prime }}$ and C$^{^{\prime }}$ are given as. 
\begin{eqnarray}
\rho _{B^{^{\prime }}} &=&Tr_{A^{^{\prime }},C^{^{\prime }}}\left( \left|
\psi ^{^{\prime }}\right\rangle \left\langle \psi ^{^{\prime }}\right|
\right) =\frac{5}{6}\left| \chi ^{^{\prime }}\right\rangle \left\langle \chi
^{^{\prime }}\right| +\frac{1}{6}\left| \chi ^{^{\prime }\bot }\right\rangle
\left\langle \chi ^{^{\prime }\perp }\right| ,  \nonumber \\
\rho _{C^{^{\prime }}} &=&Tr_{A^{^{\prime }},B^{^{\prime }}}\left( \left|
\psi ^{^{\prime }}\right\rangle \left\langle \psi ^{^{\prime }}\right|
\right) =\frac{5}{6}\left| \chi ^{^{\prime }}\right\rangle \left\langle \chi
^{^{\prime }}\right| +\frac{1}{6}\left| \chi ^{^{\prime }\bot }\right\rangle
\left\langle \chi ^{^{\prime }\perp }\right| ,
\end{eqnarray}
By checking the Eqs. (2) and (7), it is found that $\rho _{B^{^{\prime }}}$
and $\rho _{C^{^{\prime }}}$ are the completely transformed versions of $%
\rho _{B}$ and $\rho _{C}$, 
\begin{eqnarray*}
\rho _{B^{^{\prime }}} &=&Tr_{A^{^{\prime }},C^{^{\prime }}}\left( \left|
\psi ^{^{\prime }}\right\rangle \left\langle \psi ^{^{\prime }}\right|
\right) =UTr_{A,C}\left( \left| \psi \right\rangle \left\langle \psi \right|
\right) U^{\dagger }=U\rho _{B}U^{\dagger }, \\
\rho _{C^{^{\prime }}} &=&Tr_{A^{^{\prime }},B^{^{\prime }}}\left( \left|
\psi ^{^{\prime }}\right\rangle \left\langle \psi ^{^{\prime }}\right|
\right) =UTr_{A,B}\left( \left| \psi \right\rangle \left\langle \psi \right|
\right) U^{\dagger }=U\rho _{C}U^{\dagger }.
\end{eqnarray*}

Consequently, The controller achieves the task of two-object remote control
only by LOCC and pre-shared entanglement.

As stated in \cite{3}, there is no distillable pairwise entanglement in the
unlockable bound entangled state $\rho _{DEFG}^{ub}$ if no joint operations
are allowed for qubits in different locations. So the entanglement in the
bound entangled state alone cannot be enough for faithfully transmitting of
quantum information by LOCC. It is the entanglement existing in the state $%
\left| \psi \right\rangle _{ABC}$ that assists the bound entangled state for
transmitting the quantum information. The entanglement in the state $\left|
\psi \right\rangle _{ABC}$ is crucial for the remote information
concentration back to the controller. Therefore the ancillary qubit A plays
important role in the two-object remote control so that it cannot be
discarded.

As a result, our scheme only uses two particular four-party entangled state
to achieve deterministically complete quantum control on two remote objects.
The two object qubits and an acillary qubit are in an entangled state.

Our scheme is easily generalized to the case of more than two objects by
using more complex multipartite entangled states. Of course, the
entanglement in the whole initial state of the objects and ancillas is
required. The multi-object remote quantum control is important in future
quantum network. And the scheme can act as a multi-receiver quantum key
distribution if the quantum key is encoded in quantum state as operation
information. The receivers can retrieve the key by comparing the initial
sending state with the final receiving state. Since our scheme involves
complicated multipartite entangled state and many qubits, whether there is
more effective and simpler scheme of multi-objects remote quantum control is
still an open question.

This work has been financially supported by the National Natural Science
Foundation of China under the Grant No.10074072.

\end{document}